\documentclass[utf8]{frontiersFPHY} 

\setcitestyle{square} 
\usepackage{url,hyperref,lineno,microtype,subcaption,slashed}
\usepackage[onehalfspacing]{setspace}
\def\keyFont{\fontsize{8}{11}\helveticabold }
\def\firstAuthorLast{Gennaro Corcella} 
\def\Authors{Gennaro Corcella\,$^{1,*}$}
\def\Address{$^{1}$INFN, Laboratori Nazionali di Frascati,
  Via E. Fermi 40, 00044, Frascati (RM), Italy}
\def\corrAuthor{Gennaro Corcella}

\def\corrEmail{gennaro.corcella@lnf.infn.it}

\begin{document}
\onecolumn
\firstpage{1}


\title{The top-quark mass: challenges in definition and determination}

\author[\firstAuthorLast ]{\Authors} 
\address{} 
\correspondance{} 

\extraAuth{}

\maketitle

\begin{abstract}

The top-quark mass is a parameter of paramount importance in
particle physics, playing a crucial role in the electroweak
precision tests and in the stability of the Standard
Model vacuum. I will discuss
the main strategies to extract the top-quark
mass at the LHC and the interpretation
of the measurements in terms of well-posed top-mass definitions,
taking particular care about renormalon ambiguities, progress in
Monte Carlo event generators for top physics and 
theoretical uncertainties.

\section{}

\tiny
\keyFont{ \section{Keywords:} Colliders, heavy quarks, Monte Carlo
  generators, QCD calculations, Standard Model
} 
\end{abstract}

\section{Introduction}

The mass of the top quark is a fundamental parameter of the
Standard Model, since it enters in the electroweak precision tests
\cite{Baak:2014ora} and constrained the mass of the Higgs boson 
even before its discovery at the LHC.
It plays a role in Higgs inflation model
(see Refs.~\cite{Masina:2011un,Branchina:2014usa} for some recent
work on the subject), while 
the property of  the 
electroweak vacuum to lie on the boundary
between stability and metastability regimes \cite{Degrassi:2012ry}
does depend on the actual values and definitions
of top and Higgs masses used in the computation.\footnote{Strictly speaking the stability of the electroweak
  vacuum also depends on whether there is New Physics up to the
  Planck scale or not.
 Reference \cite{Degrassi:2012ry} assumes that
  the Standard Model is valid up to the Planck scale; other alternatives
  are discussed in \cite{Branchina:2014usa}.}
  Also, in the determination of the lifetime of
the Universe, undertaken in \cite{Andreassen:2017rzq},
part of the uncertainty is related to the top-quark mass.

In such calculations, one typically assumes that 
the measured top-quark mass, whose current world average reads
$m_t=[173.34\pm 0.27{\rm (stat)} \pm 0.71{\rm (syst)}]$~GeV
\cite{ATLAS:2014wva}, corresponds to the pole mass and
eventually adds errors of the order of few hundreds MeV
to account for possible deviations from this identification. 
For instance, possible changes of the central value or of the
uncertainty on $m_t$
may affect the results in \cite{Degrassi:2012ry}, to the point
of even moving the vacuum position inside the stability or
instability regions.
It is therefore of paramount importance determining $m_t$ at the LHC
with the highest possible precision, estimating reliably
all sources of uncertainty and eventually interpreting the
results in terms of field-theory mass definitions.

More generally, the top-quark mass is determined
by comparing experimental data
with theory predictions, so that the measured mass has to be identified
with the parameter $m_t$ employed in the calculations.
From the viewpoint of the techniques used in the extraction, 
one usually labels as `standard measurements' those relying
on the direct reconstruction of the top-decay products by means of the
template, matrix-element or ideogram methods, and as `alternative
measurements' the top-mass determinations which use
suitably defined observables, such as total production cross section or
peaks/endpoints of differential distributions.
It is remarkable noticing that, up to now, such
classes of mass determinations have never been combined.

From the theory side, as most top-mass extractions use
Monte Carlo shower codes,
one traditionally defines `Monte Carlo mass' the quantity which is
determined.
On the other hand, one refers to pole- or $\overline{\rm MS}$-mass
extraction whenever a measurement is compared with a
fixed-order, possibly resummed QCD calculation employing a given
field-theory mass definition. 
The distinction between Monte Carlo and well-posed mass definitions
like the pole mass has been the core of several discussions
within the top-quark physics community, as we have authors trying
to quantify the discrepancy between such masses, finding
results of the order of a few hundreds of MeV
\cite{Fleming:2007qr,Hoang:2008xm,Butenschoen:2016lpz,Hoang:2017kmk,Hoang:2018zrp}
and others who
instead present arguments against the classification of some
measurements as Monte Carlo mass determinations
\cite{Nason:2016tiy,Nason:2017cxd}
and try to interpret 
them still as pole-mass extractions, with an uncertainty which depends
on the specific measurement strategy and details of the event generation.
Furthermore, as will be discussed later,
even the so-called pole or $\overline{\rm MS}$ mass determinations
are not completely
Monte Carlo independent, since the evaluation of the
experimental acceptance
depends, though quite mildly, on the shower code which is employed
and on the implemented mass parameter.

Another issue that was often used to argue against the employment of the
pole mass has been the infrared
renormalon ambiguity \cite{Beneke:1994sw,Beneke:1994rs},
namely the factorial growth
of the coefficients of the expansion in powers of the strong coupling of 
the heavy-quark self energy, whenever it is expressed in terms of the
pole mass. However, recent work on this topic
\cite{Beneke:2016cbu,Hoang:2017btd} showed that, using
the 4-loop relation between pole and (renormalon-free)
$\overline{\rm MS}$ masses \cite{Marquard:2015qpa}, the renormalon ambiguity is actually of the order
at most of 250 MeV, hence smaller than the current error on the top mass.
Furthermore, although the projections for the future high-energy and
high-luminosity runs of the LHC aim at even lower uncertainties,
it should always be reminded that the top quark is an unstable particle
with a width of the order of 1 GeV
which, as long as it is included in the computation,
acts as a cutoff for radiation off top quarks.\footnote{The latest Particle Data Group \cite{Tanabashi:2018oca}
quotes a top width $\Gamma_t=\left(1.41^{+0.19}_{-0.15}\right)$~GeV.}

In the following, I shall give an overview of the
up-to-date top-mass determinations and, above all, I will try 
to stress the main points of the existing controversies concerning
mass definitions and interpretation of the LHC measurements,
as well as the sources of theory uncertainty.
In Section 2 I shall review the heavy-quark mass definitions; in Section
3 I will discuss the renormalon ambiguity; in Section 4 
the main strategies to measure the top mass will be presented.
The interpretation of the measurements and the theoretical uncertainties
will be investigated in Section 5, while Section 6 will contain
some final remarks.

\section{Top-quark mass definitions}

Heavy-quark mass definitions are related to how one
subtracts the ultraviolet divergences in the 
renormalized heavy-quark self energy $\Sigma^R$. 
Higher-order corrections to the self energy are typically calculated in
dimensional regularization, with
$d=4-2\epsilon$ dimensions.
At one loop in QCD, for a heavy quark with four-momentum
$p$ and bare mass $m_0$, the renormalized self energy reads:
\begin{eqnarray}\label{sigma}
\Sigma^R(m_0,p,\mu)&=&\frac{i\alpha_S}{4\pi}\left\{
\left[\frac{1}{\epsilon}-\gamma+\ln 4\pi+A(m_0,p,\mu)\right]\slashed{p}-
\left[4\left(\frac{1}{\epsilon}-\gamma+\ln 4\pi\right)+B(m_0,p,\mu)\right]m_0\right\}
\nonumber\\ &+&
i[(Z_2-1)\slashed{p}-(Z_2Z_m-1)m_0]+{\mathcal O}(\alpha_S^2),\end{eqnarray}
where $Z_2$ and
$Z_m$ are the wave-function and mass renormalization constants,
respectively, $\gamma=0.577216\dots $ the Euler--Mascheroni constant
and  $\mu$ is the renormalization scale.\footnote{In $d$ dimensions, the coupling $g_S$, related to $\alpha_S$ via $\alpha_S=g_S^2/(4\pi)$,
gets mass dimension $\epsilon$, i.e. $g_S\to g_s\mu_r^\epsilon$,
$\mu_r$ being a regularization scale.
After adding suitable counter-terms, $\Sigma^R$ is eventually
expressed in terms of the renormalization scale $\mu$.}
The functions $A$ and $B$ in Eq.~(\ref{sigma}) depend on
$p$, $m_0$ and $\mu$ and are independent of $\epsilon$.
The bare heavy-quark propagator is  $S^0(p)=i/(\slashed{p}-m_0)$,
while the renormalized $S^R$ reads,
in terms of the renormalized self energy:
\begin{equation}
  S^R(p,\mu)=\frac{i}{\slashed{p}-m_0-i\Sigma^R(m_0,p,\mu)}.
\end{equation}
The on-shell renormalization scheme, leading to the pole mass,
is defined so that the self energy and its partial derivative with respect
to $\slashed{p}$ vanish whenever $\slashed{p}=0$:
\begin{equation}
    \Sigma^R\Big |_{\slashed{p}=0}=0\ \ ;\ \ 
    \frac{\partial\Sigma^R}{\partial\slashed{p}}\Big |_{\slashed{p}=0}=0.
\end{equation}
The minimal-subtraction $(\overline{\rm MS})$ scheme
is indeed typical of dimensional regularization and 
fixes 
$Z_2$ and $Z_m$ in order to subtract just the contributions
$\sim \frac{1}{\epsilon}-\gamma+\ln 4\pi$ in Eq.~(\ref{sigma}).
\footnote{Alternatively to working in $d$ dimensions, one can 
use a mass regularization scheme, giving the gluon
a fictitious mass $\lambda$.
The renormalized self energy with a gluon mass $\lambda$ can
be obtained from (\ref{sigma}) by means of the
replacement:
  $1/\epsilon-\gamma+\ln[(4\pi\mu^2)/m_0^2]\to \ln(\lambda^2/m_0^2)$.}

Since pole and $\overline{\rm MS}$ masses are the most popular
top-mass schemes, hereafter I will devote some discussion on
such definitions.
In the on-shell (o.s.) and 
$\overline{\rm MS}$ schemes $S^R(p)$ can then be expressed in terms of pole and
$\overline{\rm MS}$ masses, respectively, as follows:
\begin{equation}\label{sr}
S^R_{\rm o.s.}(p)\simeq \frac{i}{\slashed{p}-m_{\rm pole}}\ \ ,\ \  
S^R_{\overline{\rm MS}}(p,\mu)\simeq \frac{i}{\slashed{p}-m_{\overline{\rm MS}}(\mu)-
  (A-B) m_{\overline{\rm MS}}(\mu)}.\end{equation}
From Eq.~(\ref{sr}), one can learn that 
$m_{\rm pole}$ is still the pole
of the propagator, even after the
renormalization procedure, which is in agreement with
the intuitive notion of the mass of a
free particle, whereas $m_{\overline{\rm MS}}(\mu)$
may be quite far from the pole. Also, unlike the pole mass,
the $\overline{\rm MS}$ mass depends on the renormalization scale $\mu$.
The relation between top-quark pole ($m_{t,\rm pole}$) 
and $\overline{\rm MS}$ ($\bar m_t(\bar m_t)$) masses was calculated
up to four loops in \cite{Marquard:2015qpa} and reads:
\begin{eqnarray}\label{polems}
m_{t,\rm pole}&=&\bar m_t(\bar m_t)
\left[1+0.4244~\alpha_S+0.8345~\alpha_S^2+2.375~\alpha_S^3+
  (8.615\pm 0.017)~\alpha_S^4+{\mathcal O}(\alpha_S^5)\right]\nonumber\\
&=& [163.508+7.529+1.606+0.496+(0.195\pm 0.0004)]~{\rm GeV}.
\end{eqnarray}
The last term in (\ref{polems}) yields
an uncertainty of about 200~MeV on the pole-$\overline{\rm MS}$
conversion. Beyond four loops, one can find in
Ref.~\cite{Kataev:2018mob} the dependence of the five- and
six-loop corrections to the pole-$\overline{\rm MS}$ relation 
on the number of light flavours.

  As discussed in the introduction, higher-order corrections to
the self energy, when expressed in terms 
of the pole mass, lead to infrared renormalons
\cite{Beneke:1994sw}, namely the factorial growth of
the coefficients of $\alpha_S^n$: we shall discuss recent calculations
on renormalons in the next section.
For the time being, I just point out that
the $\overline{\rm MS}$ mass is 
renormalon-free and it is therefore a so-called
short-distance mass, well defined in the infrared regime.
However, differently from the pole mass,
it is not a suitable mass definition at threshold,
as it exhibits corrections $(\alpha_S/v)^k$, 
$v$ being the top velocity, that are large in the threshold limit $v\to 0$.
On the contrary, the $\overline{\rm MS}$ mass is appropriate to describe
processes far from threshold, i.e. at scales $Q\gg m_t$ for top quarks,
since, by setting the renormalization scale $\mu\simeq Q$, one is
capable of resumming large logarithms $\ln(Q^2/m_t^2)$
in the mass definition itself.
As will be highlighted 
in the next section, Eq.~(\ref{polems}), relating the pole
mass to the renormalon-free $\overline{\rm MS}$ one, 
can be used as a starting point  
to evaluate the renormalon ambiguity in the top pole mass.


Another mass definition, which has been employed
especially in the framework of Soft Collinear Effective
Theory (SCET), is the so-called MSR mass, which was introduced
to interpolate between pole and $\overline{\rm MS}$ masses
\cite{Fleming:2007qr}.
Such a mass, labelled as $m_t^{\rm MSR}(R,\mu)$ for top quarks,
besides the renormalization scale $\mu$, depends on an extra scale
$R$, in such a way that:
\begin{equation}
m_t^{\rm MSR}(R)\to m_{t,\rm pole}\  \  {\rm for}\ \   
R\to 0\  \ {\rm and}\  \ m_t^{\rm MSR}(R)\to \bar m_t(\bar m_t)\ \ {\rm for}\ \   
R\to \bar m_t(\bar m_t) .
\end{equation}
The MSR mass can be related to any other mass definitions, such
as the pole mass, by means of a counterterm like:
\begin{equation}
  m_{t,\rm pole}=m_t^{\rm MSR}(R,\mu)+\delta m_t(R,\mu),
  \end{equation}
where the $\mu$-dependence of $m^{\rm MSR}(R,\mu)$ follows
renormalization group equations. As will be argued in the following,
the MSR mass has often been adopted in the literature to connect the
top-mass measurements with well-defined top-mass definitions,
with $R\sim {\mathcal O}(1~{\rm GeV})$.

For the sake of generality, although the present review will be mostly
devoted to hadron-collider top-mass determinations, I wish to
remind some other top
mass definitions which are often employed in analyses 
on the $m_t$ extraction at future lepton colliders.
In fact, physical observables at threshold,
such the $t\bar t$ cross section
in $e^+e^-$ collisions at 
$\sqrt{s}\simeq 2m_t$, require suitable mass schemes.
One of such definitions is the 1S mass, defined as half the mass of
a fictitious $\Upsilon(1S)$ resonance, made up of a bound $t\bar t$
state \cite{Hoang:1999zc}:
\begin{equation}
  m_{t,1{\rm S}}=\frac{1}{2}\left\{m\left[\Upsilon(1{\rm S})\right]\right\}.
  \end{equation}
The 1S mass reads, in terms of the pole mass:
\cite{Hoang:2001mm}:
\begin{equation}
  m_{t,1{\rm S}}=m_{t,{\rm pole}}\left(1-\Delta^{\rm LL}-\Delta^{\rm NLL}-
    \Delta^{\rm NNLL}\right).
  \end{equation}
The explicit expression of the $\Delta$ terms can be found
in \cite{Hoang:2001mm}, where the threshold $e^+e^-\to t\bar t$
cross section was computed in the next-to-next-to-leading
logarithmic approximation,
and the superscripts LL, NLL and NNLL
refer to the resummation of large logarithms of the top velocity $v$,
which are large in the regime $v\sim\alpha_S\ll 1$ and $\alpha_S\ln v\sim 1$.

The potential-subtracted (PS) mass is instead constructed in terms of 
the $t\bar t$ Coulomb potential, in such a way that 
contributions below a factorization scale 
$\mu_F$ are subtracted off, as to suppress renormalons \cite{Beneke:1998rk}:
\begin{equation}\label{mps}
m_{\mathrm{PS}}(\mu_F)=m_{\mathrm{pole}}-
\frac{1}{2}\int_{|q|<\mu_F}{\frac{d^3 q}{(2\pi)^3}}
\tilde V(q).
\end{equation}
In Eq.~(\ref{mps}) $\tilde V(q)$ is the transform in momentum space of the
$t\bar t$ Coulomb potential.
The PS mass is a threshold mass too, particularly
suitable to deal with $t\bar t$ production at energies slightly
above $2m_t$.
The relation 
between PS and pole top-quark masses is given by the following equation
\cite{Hoang:2000yr}:
\begin{equation}
  m_{t,\rm PS}(\mu_F)=m_{t,{\rm pole}}-\frac{4}{3\pi}\alpha_S(\mu_F)\mu_F+{\mathcal O}(\alpha_S^2).
  \end{equation}
More recently, the theoretical error on the possible extraction
of 1S and PS masses in $e^+e^-$ collisions just above the
$t\bar t$ threshold was estimated.
In detail, by using a NNLL threshold
resummation of the ratio $R=\sigma(e^+e^-\to t\bar t)/
\sigma(e^+e^-\to \mu^+\mu^-)$, the 1S mass can be extracted
with an uncertainty about 40 MeV \cite{Hoang:2013uda}, whereas, by employing a
fixed-order NNNLO
calculation, the PS mass can be determined with an error below
50 MeV \cite{Beneke:2015kwa}.
It will be of course desirable to combine such fixed-order
and resummed computations to possibly decrease further such an
uncertainty.

Another threshold mass definition is the renormalon-subtracted (RS) mass, which 
removes from the pole mass the pure renormalon contribution
\cite{Pineda:2001zq}. The RS mass was determined in 
\cite{Pineda:2001zq} after constructing its Borel transform and
reads, in terms of the pole mass:
\begin{equation}
  m_{t,\rm RS}=
  m_{t,\rm pole}-\sum_{n=0}^\infty N_m\mu_F
  \alpha_S^{n+1}(\mu_F)\sum_{k=0}^\infty c_k
  \frac{\Gamma(n+1+b-k)}{\Gamma(1+b-k)},\end{equation}
where the expression for the coefficients
$N_m$ and $c_k$ are given in \cite{Pineda:2001zq}
and $b$ can be expressed in terms of the QCD $\beta$-function
as $b=\beta_1/(2\beta_0^2)$.
Potential-, renormalon-subtracted and 1S 
top-quark masses were related to the
$\overline{\rm MS}$ mass in Ref.~\cite{Marquard:2015qpa}
with four-loop accuracy in the conversion.
The uncertainty in the conversion was gauged about
7, 11 and 23 11 MeV for PS, RS and 1S masses, respectively.

Finally, the so-called kinetic mass was defined in
\cite{Bigi:1996si} for the purpose of
improving the convergence of the perturbative expansion of the
semileptonic $B$-meson decay width. It was constructed
by subtracting from the pole mass the
HQET (Heavy Quark Effective Theory)
matrix elements, denoted by
$\bar\Lambda(\mu)$ in \cite{Bigi:1996si}, expressing the
shift between pole and meson masses.
The kinetic bottom-quark mass reads,
up to terms suppressed as the inverse of the quark/meson mass:
\begin{equation}
  m_{b,\rm kin}(\mu_F)=m_B-\bar\Lambda (\mu_F)+{\mathcal O}\left(\frac{1}{m_B}\right).
  \end{equation}
In \cite{Hoang:2000yr}, the kinetic mass was generalized to
$t\bar t$ bound states, obtaining the following expansion in
terms of the pole mass:
\begin{equation}
  m_{t,{\rm kin}}(\mu_F)=m_{t,{\rm pole}}-
  \frac{16}{9\pi}\alpha_S(\mu_F)\mu_F+
       {\mathcal O}(\alpha_S^2).
       \end{equation}
As underlined before, the 1S, PS and RS masses are threshold
masses which, unlike the
pole mass, do not exhibit the renormalon ambiguity.
Recent calculations aimed at estimating the renormalon
uncertainty in the pole mass will be the topic of next
section.

\section{The renormalon ambiguity in the top mass}
Problems with the renormalized
heavy-quark self energy, when expressed in terms of the pole mass,
were first
understood in \cite{Beneke:1994sw,Beneke:1994rs}.
In fact, after including higher-order contributions
in the strong coupling constant, the renormalized 
heavy-quark self energy exhibits the following expansion in powers
of $\alpha_S$:
\begin{equation}\label{fact}
\Sigma^R(m_{\rm pole},m_{\rm pole})\approx m_{\rm pole}\ \sum_n\ \alpha_S^{n+1}\  
(2b_0)^n\ n!,
\end{equation}
where $b_0$ is first $\beta$-function coefficient
entering in the $\overline{\rm MS}$ strong coupling constant.\footnote{We recall that,
  e.g., at LO in the  $\overline{\rm MS}$ scheme,
  it is $\alpha_S(Q^2)=1/[b_0\ln(Q^2/\Lambda^2)]$, $\Lambda$ being
  the QCD scale.
  For $Q^2\sim\Lambda^2$ one hits the well-known Landau pole
and perturbative QCD can no longer be applied.}
From Eq.~(\ref{fact}), one learns that the coefficients of
the expansion grow like $n!$ at order $\alpha_S^{n+1}$.

After re-expressing $\alpha_S$ in terms of
the $\beta$ function and of the QCD scale $\Lambda$, and
inserting $\Sigma^R$ in the on-shell propagator (\ref{sr}),
one will get 
a correction to the pole mass:
\begin{equation}
\Delta m_{\rm pole}\simeq {\mathcal O}(\Lambda),
\end{equation}
which is the renowned renormalon ambiguity in $m_{\rm pole}$,
i.e. an uncertainty of the order of the QCD scale in the pole-mass
definition.
This result can be related to the fact that a quark is not
a free parton, but has to be confined into a hadron:
in fact, one can prove that the renormalon uncertainty is due to
the gluon self coupling, while it is not present when dealing
with leptons. Therefore, the pole mass behaves like
a physical mass for electrons or muons, whereas for heavy quarks
it is not a short-distance mass, because of infrared
renormalon effects, and one should choose on a case-by-case basis 
whether the pole mass or other definitions are adequate to
describe a given physical process.

In order to quantify the renormalon ambiguity in the
pole mass, one can employ the relation between pole and
$\overline{\rm MS}$ masses, relying on the fact that the
$\overline{\rm MS}$ mass is unaffected by renormalons.
Equation (\ref{polems}) can be parametrized to all orders as
in Ref.~\cite{Beneke:2016cbu}:
\begin{equation}\label{asym}
  m_{\rm pole}=\bar m(\mu_m)\left[1+\sum_{n=1}^\infty c_n(\mu, \mu_m,\bar m(\mu_m))
    \alpha_S^n(\mu)\right],
  \end{equation}
with $\bar m(\mu_m)$ being the $\overline{\rm MS}$ mass at some
scale $\mu_m$ and $\mu$ the renormalization scale at which
the strong coupling is evaluated.
The dominant renormalon divergence implies that the coefficients
$c_n$ in the asymptotic expansion
have to satisfy the following relation at large $n$:
\begin{equation}\label{cn}
  c_n(\mu,\mu_m,m(\mu_m))\to N\frac{\mu}{m(\mu_m)}\ c_n^{\rm as}
  \ \ {\rm for} \ \ n\to \infty.
  \end{equation}
The expression for the asymptotic coefficients
$c_n^{\rm as}$ can be found in \cite{Beneke:2016cbu} and
is consistent with the fact that the renormalon factorial
growth is due to the low-momentum region in the higher-order loop
corrections to the heavy-quark self energy.
The calculation of the normalization coefficient $N$ is non trivial:
in \cite{Beneke:2016cbu} $N$ was extracted after fitting the
third- and fourth-order coefficient in the exact four-loop
$\overline{\rm MS}$-pole mass conversion and
amounts to
$N\simeq 0.976\dots$ for $N_C=3$ number of colours.

Furthermore, an alternative and possibly better method to deal
with factorially divergent series consists in using the
Borel transform, which, for a function $f(\alpha_S)$ reads:
\begin{equation}
  f(\alpha_S)=\sum_{n=0}^\infty c_n\alpha_S^{n+1}\ ;\
  B[f](t)=\sum_{n=0}^\infty c_n\frac{t^n}{n!},\end{equation}
which implies
\begin{equation}\label{borel}
  f(\alpha_S)=\int_0^\infty{e^{-t/\alpha_S}B[f](t)}.
  \end{equation}
The evaluation of the Borel integral (\ref{borel}) depends on a prescription:
one typically takes its principal value and, following the
so-called `Im/Pi' method, the uncertainty is estimated as the modulus of
the imaginary part, arising from the integration above and below
the singular cuts in the complex plane, divided by $\pi$.
In fact, in Ref.~\cite{Beneke:2016cbu} the asymptotic expansion of
the pole mass with respect to the $\overline{\rm MS}$ one was
computed as an inverse Borel transform, by using the Im/Pi method
for the error, considering only three light flavours
and accounting for charm and bottom masses. 
The final result is that the leading renormalon ambiguity
amounts to about
110 MeV for top as well as bottom and charm pole masses.

A different strategy to gauge the renormalon ambiguity was instead
tackled in \cite{Hoang:2017btd}, where the MSR mass $m_{\rm MSR}(R)$
was used. In the relation between $m_{\rm pole}$ and $m_{\rm MSR}(R)$,
\begin{equation}\label{msrpole}
m_{\rm pole}=m_{\rm MSR}(R)+R\sum_{n=1}^\infty a_n
\left[\frac{\alpha_S(R)}{4\pi}\right]^n,
\end{equation}
the scale $R$ is set to the $\overline{\rm MS}$ top mass
$m_t(\bar m_t)$ and 
the series (\ref{msrpole}) is truncated at some fixed order
$n$. A value $n_{\rm min}$ is determined in such a way to
minimize the difference $\Delta(n)=m_{\rm pole}(n)-m_{\rm pole}(n-1)$
and a number $f$ slightly above unity is defined.
The set $\{n\}_f$ is thus constructed in such a way that
$\Delta (n)\leq f\Delta (n_{\rm min})$: the midpoint of
$m_{\rm pole}(n)$ within $\{n\}_f$ is then chosen as the central value
and half of the variation range of $m_{\rm pole}(n)$ as an estimate
of the ambiguity, accounting for the running of the renormalization
scale as well. After observing that the results
depend on $f$ rather mildly, in \cite{Hoang:2017btd} $f=5/4$ was chosen,
yielding an ambiguity about 253 MeV in the pole mass.
Both in Ref.~\cite{Beneke:2016cbu} and Ref.~\cite{Hoang:2017btd},
some thorough discussion is devoted to the inclusion of
charm and bottom masses. The results of 110 and 253 MeV would go down to
70 \cite{Beneke:2016cbu} and 180 \cite{Hoang:2017btd} MeV if one
treated charm and bottom quarks as massless.
Some attempts to relate the different methods adopted in 
\cite{Beneke:2016cbu} and \cite{Hoang:2017btd} were made in
\cite{Nason:2017cxd}. In fact, the result in \cite{Beneke:2016cbu}
can be obtained even following the method in \cite{Hoang:2017btd},
but taking as central value half the sum of all $\Delta(n)$ and
setting $f=1+1/(4\pi)$ in the uncertainty evaluation.

In the following, no strong statement 
supporting the calculation in \cite{Beneke:2016cbu} or
 \cite{Hoang:2017btd} will be made.
 I just wish to point out that,
 on the one hand, as long as the uncertainties in the
top-mass measurement
stay around 500 GeV, both renormalon determinations are smaller
and should not play any role in supporting the use of a given mass
definition. This may not be the case if, in future perspective,
one ideally aims at precisions about 200-300~MeV.
However, as will be underlined when dealing 
with Monte Carlo modelling and theoretical errors, recent
implementations of top production and decay in shower codes include
width effects \cite{Jezo:2016ujg}, 
in such a way that the top width, about 1.4 GeV
and well above the energy range of both renormalon estimates, 
acts as a cutoff for the
radiation off top quarks. 
\footnote{This would not be the case in codes or calculations 
  which instead neglect width
  effects and interference between top-production and decay phases.
  In this case, even top quarks are capable of radiating down to the
infrared cutoff.}
Of course, if one considers observables relying on top decays
($t\to bW$), the $b$-quark is allowed to emit soft radiation down
to the shower cutoff and, in principle, in quantities depending
on $b$-jets one may have to deal with renormalons.

A careful exploration of renormalon effects 
in observables depending on the
top mass was carried out in
Ref.~\cite{FerrarioRavasio:2018ubr}. The authors found that the
$\overline{\rm MS}$ mass is a better definition
for quantities like the total $t\bar t$ 
cross section, while using the pole mass would lead to
a linear renormalon and an ambiguity of ${\mathcal O}(100~{\rm MeV})$
on the $m_t$ extraction. 
Indications in favour of such a short-distance mass
were also given whenever final-state jets are reconstructed
using algorithms with a large jet radius $R$.
As for the reconstructed top mass from, e.g., the $b$-jet+$W$ invariant
mass, in the pole-mass scheme 
a linear renormalon correction is present, whose coefficient is
nevertheless pretty small if one employs
a large $R$ in the $b$-jet definition.
Finally, leptonic observables exhibit a linear renormalon with both
mass definitions, as long as one works in the narrow-width approximation.
On the contrary, there are no linear renormalons if one adopts a
a short-distance mass and includes the finite top width.

\section{Top-quark mass extraction at LHC}

Top-quark mass determinations at hadron colliders
are classified
as standard or
alternative measurements and, according to the decay modes of the
two $W$'s in top decays, as measurements in the dilepton, lepton+jet
or all-hadronic channels. 
Standard top-mass analyses are based on the direct reconstruction
of top-decay final states and 
compare observables, such as the $b$-jet+lepton invariant-
mass distribution,
with the predictions yielded by
the Monte Carlo codes. So-called alternative measurements 
use instead other observables, such as total/differential cross sections 
or distribution peaks/endpoints.
Since, as will be detailed in the following, Monte Carlo codes are of
paramount importance for most top-mass analyses, I shall first sketch their
main features, and then review the experimental methods to extract $m_t$.

\subsection{Monte Carlo generators for top physics}

The last couple of decades has seen a tremendous progress 
in the implementation of Monte Carlo event generators, besides the
reknowned general-purpose HERWIG \cite{Corcella:2000bw,Bellm:2015jjp}
and PYTHIA \cite{Sjostrand:2006za,Sjostrand:2014zea}, in such a way that
several reliable programs are currently available for
the top-mass analyses.
On the one hand, strategies to match NLO calculations with
parton showers were developed, on the other one a number
of so-called matrix-element generators were released.
In fact, matrix-element generators simulate multi-leg amplitudes and
are interfaced to HERWIG or PYTHIA for shower and hadronization:
besides top-quark signals,
they are very useful to simulate backgrounds
with high jet multiplicities, such as $W/Z+n$~jets, which
would be poorly described by HERWIG or PYTHIA for $n>1$.

Regarding top phenomenology, standard Monte Carlo programs
like 
simulate both top production and decays using leading order (LO)
matrix elements,
multi-parton emission in the soft or collinear
limit and the interference between
top-production and decay stages is neglected (narrow-width
approximation). HERWIG parton showers satisfy angular ordering
\cite{Marchesini:1983bm,Gieseke:2003rz}, with the latest version
even allowing the option of dipole-like evolution \cite{Platzer:2011bc};
PYTHIA cascades
  are instead ordered in transverse momentum.
  \footnote{The old PYTHIA 6 code also implements virtuality ordering,
    with the option to veto non-angular-ordered emissions.}
Matrix-element corrections to parton showers
are implemented for top decays \cite{Corcella:1998rs,Norrbin:2000uu},
but not for production,
and the total production cross section and top-decay width
are still calculated at LO.
Hadronization is included by adopting the
cluster model \cite{Webber:1983if}, based on colour pre-confinement,
in HERWIG and the string model \cite{Andersson:1983ia}
in PYTHIA.
The underlying event used to be described assuming
soft collisions between the proton spectators and tuning
the model parameters to minimum-bias events at small transverse momentum.
Nevertheless, all modern codes implement it
through multiple scatterings
strongly ordered in transverse momentum:
the underlying event is thus a secondary collision, whose transverse
momentum is much lower than the primary hard scattering
\cite{Butterworth:1996zw,Sjostrand:2004pf}.

Among the new generation of Monte Carlo programs,
SHERPA \cite{Gleisberg:2008ta} can also be considered a multi-purpose
code, in the light of the wide spectrum of processes which it
is capable of simulating.
In detail, matrix elements are
computed by means of the AMEGIC++ \cite{Krauss:2001iv}
and COMIX \cite{Gleisberg:2008fv} codes, while the interface to
one-loop generators, implemented along the lines of
\cite{Binoth:2010xt}, allows one to include NLO QCD and
possibly electroweak corrections. Parton showers are then
accounted for according to the dipole formalism developed in
\cite{Catani:1996jh}, underlying event and hadronization
follow the multiple-scattering and cluster models in
PYTHIA and HERWIG, respectively.

For the purpose of the matching of NLO matrix elements and
multi-parton cascades,
NLO+shower programs, such as 
MadGraph5$\_$aMC@NLO \cite{Frixione:2002ik,Alwall:2014hca}
and POWHEG \cite{Alioli:2010xd}, 
implement NLO
hard-scattering amplitudes, but still depend on 
HERWIG and PYTHIA for parton cascades and non-perturbative
phenomena.
The earlier versions of such NLO+shower algorithms only
included NLO corrections to $t\bar t$ production, while (LO) top
decays and hadronization were still handled in the parton shower
approximation.
The later implementation of POWHEG
\cite{Jezo:2016ujg} includes
in the $b\bar b 4\ell$ code
both top production and decay at NLO, accounting for
the interference between
top production and decay stages, as well as non-resonant contributions
leading to $(W^+b)(W^-\bar b)$ final states.
\footnote{See also Ref.~\cite{Heinrich:2017bqp}
for an independent investigation of NLO and
top-width effects on the top-mass determination.}
As for MadGraph5$\_$aMC@NLO, strictly speaking, top decays are still
at LO, however spin correlations are included through 
the MadSpin package \cite{Artoisenet:2012st}
and, as discussed in \cite{Frederix:2016rdc}, 
they account for a significant part of the NLO corrections.
For the purpose of HERWIG, it has its own implmentation of
NLO+shower merging/matching \cite{Platzer:2012bs,Bellm:2017ktr},
working for top-quark production and decay in the narrow-width approximation
\cite{Cormier:2018tog}.

Regarding matrix-element generators, suitable codes to 
describe top-quark signals and backgrounds are, 
among others, 
ALPGEN \cite{Mangano:2002ea}, MCFM
\cite{Campbell:2010ff}, CalcHEP
\cite{Belyaev:2012qa}, HELAC \cite{Kanaki:2000ey}
and WHIZARD \cite{Kilian:2007gr}.
In particular, ALPGEN and  CalcHEP
simulate multi-parton final states at LO and 
can be interfaced to HERWIG or PYTHIA for shower and hadronization.
HELAC and WHIZARD have been lately provided with NLO corrections
\cite{Bevilacqua:2011xh,Weiss:2015npa} and matching to 
shower and hadronization codes as well.
MCFM is a NLO parton-level Monte Carlo code: top production and
decay are handled at NLO, in the narrow-width approximation.

Before concluding this subsection, it is worthwhile saying a few
words on the precision of the predictions yielded by
Monte Carlo codes.
As observed before, parton showers simulate multiple radiation in
the soft or collinear approximation and, in general, the accuracy of a
prediction depends on the specific observable under investigation.
Although total cross sections and widths are (N)LO, 
for most quantities
Monte Carlo predictions are equivalent to leading-logarithmic
resummations, i.e. they resum double soft and collinear logarithms,
and include some classes of subleading logarithms, i.e. only
soft- or collinear-enhanced.\footnote{A notable exception
  is given by leading
  non-global logarithms, sensitive to a limited portion of the
  phase space, which, as discussed in \cite{Banfi:2006gy}, are
  partially accounted for by the angular-ordered showers of
  HERWIG, while they are mostly absent in virtuality- or
  transverse-momentum-ordered PYTHIA.}
Reference~\cite{Catani:1990rr} even proved that, in Deep Inelastic Scattering
and Drell--Yan processes at large values of the Bjorken $x$,
the HERWIG algorithm is capable of capturing all next-to-leading
logarithms, i.e. all single logarithms, enhanced for soft or
collinear emission, as long as one rescales the QCD scale
$\Lambda$ to a Monte Carlo value, labelled $\Lambda_{\rm MC}$.
\footnote{With respect to $\Lambda$ in the $\overline{\rm MS}$ scheme,
  it is $\Lambda_{\rm MC}=\Lambda\exp[K/(4\pi b_0)]$,
  with $K=N_C(67/18-\pi^2/6)-5N_f/9$, with $N_C$ and $N_f$ being the
  number of colours and active flavours, respectively.}

\subsection{Standard and alternative top-mass measurements}

In this subsection I shall briefly present the main strategies
to measure the top mass at hadron colliders in $t\bar t$ events,
taking particular care about the analyses carried out at the LHC.

\subsubsection{Direct reconstruction methods}
  
\begin{figure}[ht!]
\centering\vspace{-0.5cm}\hspace{-3.cm}
\includegraphics[width=0.8\textwidth]{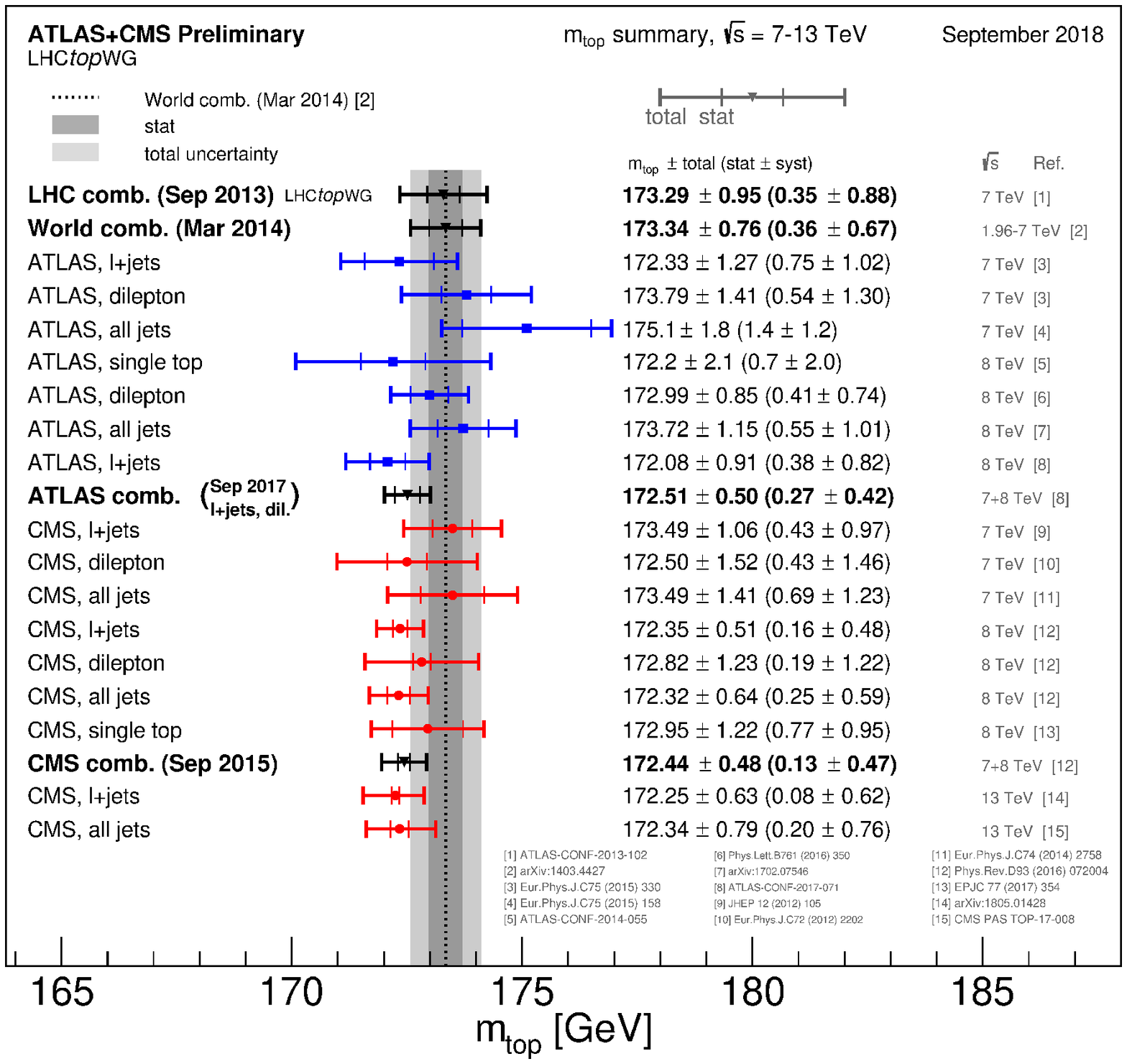}\vspace{-1.5cm}
\caption{Summary of the top-mass analyses at the LHC, accounting
  for the world average and the ATLAS and CMS combinations as well.}
\label{summ}
\end{figure}

Strategies based on the direct reconstruction
of the top-decay products, namely the template,
matrix-element and ideogram methods, 
have been traditionally classified as
standard top-mass determinations. 
As for ATLAS, the most up-to-dated measurements 
are given at 8 TeV and 19.7~fb$^{-1}$ in
Refs.~\cite{Aaboud:2016igd,Aaboud:2018zbu,Aaboud:2017mae}
for dilepton, lepton+jets and all-hadronic modes, respectively.
Regarding CMS, at the moment even results at $\sqrt{s}=13$~TeV
and ${\mathcal L}=35.9$~fb$^{-1}$ are available and are reported
in \cite{Sirunyan:2018goh} (dileptons),
\cite{Sirunyan:2018gqx} (lepton+jets) and 
\cite{Sirunyan:2018mlv} (all hadrons). 
Regarding these analyses and summing in quadrature systematic
and statistical errors, 
CMS quotes uncertainties
about 0.73 GeV for dileptons, 0.62 GeV for leptons+jets
and 0.61 GeV for the all-hadron channel.
As for ATLAS, the uncertainties are 0.84 GeV (dileptons),
0.91 GeV (lepton+jets) and 0.73 GeV (all jets).
The standard top-mass measurements have been the basis to determine
the world average \cite{ATLAS:2014wva}, already presented
in the introduction, which, after summing statistical and systematic
errors in quadrature, yields an overall uncertainty 
about 800 MeV.
Work towards an updated world average is currently under way.
The LHC collaborations 
have nevertheless released their own combined measurements
using 7 and 8 TeV data together: details on such studies
can be found in Refs.~\cite{Aaboud:2018zbu,Khachatryan:2015hba}
for ATLAS and CMS, respectively. 
Both analyses yield a total error about 0.5 GeV, hence an overall
precision on the top mass around 0.3\%. Figure~\ref{summ}
summarizes the state of the art on top-mass measurements carried out at the
LHC, including the world average, as well as ATLAS and CMS combinations.

As discussed in the introduction, 
since the standard $m_t$-reconstruction
methods rely on the use of Monte
Carlo generators, such measurements are usually 
quoted as `Monte Carlo mass' and 
much debate has been taking place on whether
the extracted mass can be related to any well-posed definition,
with some calculable uncertainty,
such as the pole mass.
The ongoing discussion on the theoretical interpretation
of the measured top mass
will be the main topic of next session.
Before moving to this issue, it is worthwhile reviewing
the so-called `alternative' strategies, making use of total/differential
cross sections, endpoints or other kinematic
properties of $t\bar t$ final states.

\subsubsection{Total and differential $t\bar t$ 
  cross section}

The total $t\bar t$
cross section was calculated in QCD in 
the NNLO+NNLL approximation
in Ref.~\cite{Czakon:2013goa}
\footnote{At NNLO the $t\bar t$ cross section is ${\mathcal O}(\alpha_S^4)$,
  whereas the threshold logarithms which are resummed in
  \cite{Czakon:2013goa} are $\sim\alpha_S^n[\ln^m(1-z)/(1-z)]_+$,
  with $z=m_t^2/\hat s$, $\hat s$ being the partonic centre-of-mass energy
  and $m\leq 2n-1$.}
and was used to determine $m_t$ by ATLAS in 
Refs.~\cite{Aad:2014kva} (7 and 8 TeV data) and by CMS in
\cite{Khachatryan:2016mqs} (7 and 8 TeV)
and \cite{Sirunyan:2017uhy} (13 TeV).
Since the calculation in \cite{Czakon:2013goa} employed the pole mass
definition, the results in Refs.~\cite{Khachatryan:2015hba,Aad:2014kva,
  Sirunyan:2017uhy} are quoted as pole mass measurements.
Although to some extent this is mostly correct, it should always be reminded
that even those analyses are not completely independent of the
shower generator, and therefore of its mass parameter, which is still
used to evaluate the acceptance. Nevertheless, it was proved that
such a sensitivity is rather mild.
Overall, the errors in \cite{Aad:2014kva,Khachatryan:2016mqs,Sirunyan:2017uhy}
are larger than those in the standard methods, as they are about 2.5 GeV;
however, they are expected to decrease thanks to the higher
statistics foreseen in the LHC future runs.
After the computation of the total cross section, even differential
distributions were calculated at NNLO in \cite{Czakon:2015owf},
still using the top pole mass:
this computation was used by the D0 Collaboration
\cite{Czakon:2016teh} to extract the top mass
at the Tevatron accelerator, namely $\sqrt{s}=1.96$~TeV and
${\mathcal L}=9.7~{\rm fb}^{-1}$. The error on this
measurement is about 2.5~GeV, hence competitive with those obtained
at the LHC from the total production cross section.

Reference~\cite{Dowling:2013baa} explored the extraction of
the top mass by using the
NNLO total $t\bar t$ cross section and NLO differential distributions,
such as transverse momentum, rapidity and $t\bar t$ invariant mass,
expressed in terms of pole and $\overline{\rm MS}$ masses.
Overall, Ref.~\cite{Dowling:2013baa} found that using the running mass
yields a milder scale dependence of such observables; nevertheless,
implementing the full NNLO differential cross section
or the four-loop pole-$\overline{\rm MS}$ mass conversion,
along the lines of \cite{Czakon:2015owf} and \cite{Marquard:2015qpa},
respectively,
will be obviously very useful to shed light on the scale dependence.

Still on the $t\bar t$ total cross section,
it is worthwhile pointing out the recent work carried out to
merge NNLO QCD and NLO electroweak corrections in
Ref.~\cite{Czakon:2017wor}.
Such a computation was then used to predict the top-quark
charge asymmetry at Tevatron and LHC and 
the  electroweak corrections exhibited a remarkable impact, say
about 20\%, on the forward-backward asymmetry.
It will be clearly
very interesting determining the top pole mass from
differential distributions, along the lines of \cite{Czakon:2016teh},
including electroweak contributions as well.

\subsubsection{$t\bar tj$ cross section}

The top mass was also extracted from the
measurement of the $t\bar t+1$~jet cross
section, which has a stronger sensitive to
$m_t$ than the inclusive $t\bar t$ rate.
In Ref.~\cite{Alioli:2013mxa}, the NLO $t\bar tj$ cross section was
calculated using POWHEG and its pole mass implementation,
matched to PYTHIA.
Detector and shower/hadronization effects were unfolded in order
to recover the pure NLO $t\bar tj$ cross section.
From the experimental viewpoint, the approach proposed in
\cite{Alioli:2013mxa} was followed in
\cite{Aad:2015waa} by ATLAS (7 TeV and 5 fb$^{-1}$) and
by CMS in \cite{CMS:2016khu} (8 TeV and 19.7 fb$^{-1}$).
The error on $m_t$ extracted from the $t\bar tj$ cross section is slightly
smaller than
from the inclusive $t\bar t$ one, but still much above the direct-reconstruction
measurements. Such mass determinations are referred to
as pole mass measurements, since this is the mass definition employed by
POWHEG, while the
PYTHIA mass parameter used in the parton shower has a mild effect
in the determination of the acceptance.
Reference~\cite{Fuster:2017rev} 
used the running $\overline{\rm MS}$ top mass in the calculation of 
the NLO $t\bar tj$ rate and, after comparing with the cross section
measurements, obtained results which are, within the errors,
in agreement with the pole mass yielded by the approach in 
\cite{Alioli:2013mxa}.

Other so-called alternative methods to reconstruct $m_t$ 
rely on the kinematic properties of
top-decay final states: since they are based on 
 the comparison with Monte Carlo predictions,
the measured
$m_t$ has to be identified with the mass parameter in the shower code.
Overall, such techniques yield uncertainties in the mass about the
order of magnitude of those relying on the total cross section, say
about 1 GeV or above.

\subsubsection{Peak of the $b$-jet energy spectrum}

It was observed that the peak of the $b$-jet energy
in top decay at LO is independent of the boost 
from the top to the laboratory frame, as well as of the
production mechanism \cite{Agashe:2016bok}.
The CMS Collaboration did measure the top mass from the $b$-jet
energy peak data at 8 TeV and 19.7 fb$^{-1}$ \cite{CMS:2015jwa},
by using POWHEG and MadGraph to simulate top production and decay, and
PYTHIA for parton shower, hadronization and underlying event.
The resulting uncertainties are 1.17 GeV (statistics) and
2.66 GeV (systematics).

\subsubsection{$m_{b\ell}$, $m_{b\ell\nu}$ and
  stranverse mass $m_{T2}$}

The $b$-jet+lepton invariant-mass
($m_{b\ell}$) spectrum was used by CMS to reconstruct 
$m_t$ in the dilepton channel in Ref.~\cite{CMS:2014cza},
at 8 TeV and 19.7~fb$^{-1}$. The data were compared with the
MadGraph+PYTHIA simulation,
yielding a measurement consistent with the world average and
an uncertainty about 1.3 GeV. In Ref.~\cite{CMS:2014cza},
for the sake of comparison, even the NLO code MCFM 
was used to predict the $m_{b\ell}$ distribution.
More recently, in Ref.~\cite{Sirunyan:2017idq}
CMS extracted $m_t$ even
from the so-called stransverse mass $m_{T2}$
\cite{Lester:1999tx} and from $m_{b\ell\nu}$, which
accounts for the neutrino missing transverse momentum as well.
The sensitivity of these observables to $m_t$ yields an
uncertainty about 180 MeV (statistics) and 900 MeV (systematics).

\subsubsection{Endpoint method}

Another method to measure $m_t$ consists of using the endpoints of
distributions sensitive to $m_t$, namely the endpoints
of $m_{b\ell}$, $\mu_{bb}$ and $\mu_{\ell\ell}$,
where $b$ is a $b$-flavoured
jet, and $\mu_{bb}$ and $\mu_{\ell\ell}$
generalizations of the $b\bar b$ and $\ell^+\ell^-$ invariant masses
in the dilepton channel, as described in 
Ref.~\cite{Chatrchyan:2013boa} (CMS, 7 TeV and 5 fb$^{-1}$).
Since $b$-flavoured jets can be calibrated directly
from data, the endpoint strategy is claimed to minimize the
Monte Carlo error on $m_t$, which is mostly due
to colour reconnection, namely the formation of a $B$ hadron by
combining a $b$ quark in $t$ decay with an antiquark from
$\bar t$ decay or initial-state radiation. 
Constraining the neutrino and $W$ masses to their world-average values, 
this method leads to uncertainties about 900 MeV (statistics)
and 2 GeV (systematics).

\subsubsection{Leptonic observables}

Purely leptonic observables in the dilepton channel,
such as the Mellin moments of lepton energies or transverse
momenta, 
were proposed to measure $m_t$, since in this way one can escape the
actual reconstruction of the top quarks \cite{Frixione:2014ala}.
However, this method still yields uncertainties due to 
hadronization,
production mechanism, Lorentz boost from the top
to the laboratory frame, as well as 
missing higher-order corrections.
Preliminary analyses have been carried out in \cite{CMS:2016xfv}
(CMS, based on LO MadGraph)
and \cite{Aaboud:2017ujq} (ATLAS, 
based on the MCFM NLO parton-level code
\cite{Campbell:2015qma}) using data at 8 TeV and 19.7 fb$^{-1}$ 
and are expected to be improved by
matching NLO amplitudes with shower/hadronization
generators. For the time being, the uncertainties quoted in
Ref.~\cite{CMS:2016xfv} are 1.1 GeV (statistics),
0.5 GeV (experimental systematics) and
2.5-3.1 GeV (theoretical systematics),
whereas in Ref.~\cite{Aaboud:2017ujq}
they read 0.9 GeV, 0.8 GeV and 1.2 GeV, respectively.
Reference~\cite{CMS:2016xfv} also quotes an uncertainty
$^{+0.8}_{-0.0}~{\rm GeV}$ due to the description of the top-quark
transverse momentum. In fact, previous CMS analyses had
displayed a mismodelling of the top $p_T$ simulated by
MadGraph+PYTHIA, and therefore Ref.~\cite{CMS:2016xfv} reweighted
the transverse momentum to match the measured one.

\subsubsection{$J/\psi$ method}

Final states with $J/\psi$ mesons
were exploited by the CMS Collaboration in Ref.~\cite{Khachatryan:2016pek}
to measure $m_t$, 
using data collected at 8 TeV and a luminosity about 19.7 fb$^{-1}$.
In this work, one explores $t\to bW$ processes where
$b$-flavoured hadrons decay into states containing a $J/\psi$,
the $J/\psi$
decays according to $J/\psi\to\mu^+\mu^-$ pair and 
the $W$ bosons undergo the leptonic transition $W\to\ell\nu$.
The top mass is then extracted by fitting the invariant mass
distributions $m_{\mu\mu}$ or $m_{J/\psi \ell}$, as well as the transverse momentum
of the $J/\psi$. The analysis was carried out by using
the MadGraph code, interfaced with PYTHIA, while, for the sake
of estimating the theoretical error, 
POWHEG and SHERPA were employed as well. 
Overall, the statistical uncertainty in the investigation 
\cite{Khachatryan:2016pek} amounts to 3 GeV, while the systematic error to
0.9 GeV. The conclusion of \cite{Khachatryan:2016pek}  is that,
since the systematic uncertainties are of different origin
from those entering in the measurements based on direct
reconstruction and given the higher statistics which are foreseen,
the $J/\psi$ method should ultimately be worth to be included in
the combination with the extractions from matrix-element or template
strategies. 

\subsubsection{Final-state charged particles}

A novel technique was presented by the CMS Collaboration in
Ref.~\cite{Khachatryan:2016wqo}, where $m_t$ is measured by exploiting
the kinematic properties of final-state charged particles.
The observable used in this analysis
is the mass $m_{sv\ell}$ of the secondary vertex-lepton system,
namely the invariant mass of a system made of the charged lepton in
$W$ decays and charged hadrons in a jet originating from a common secondary
vertex. Using only charged particles, in fact, reduces the overall
acceptance uncertainty, whereas this method is obviously dependent
on the modelling of top decays and bottom hadronization.
The investigation was undertaken using MadGraph+PYTHIA to simulate
the signal,  POWHEG and SHERPA to estimate the uncertainty
due to the matrix-element generation and hadronization, respectively. 
The final error on the measurement of $m_t$ from charged particles is then
200 MeV (statistics) and $^{+1.58}_{-0.97}$~GeV (systematics), by using 
data sets of 8 TeV collisions and a luminosity of 19.7 fb$^{-1}$.

\subsubsection{Perspectives at high luminosity}

\begin{figure}[t]
\centering
\vspace{-0.5cm}\hspace{-3.cm}
\includegraphics[width=0.8\textwidth]{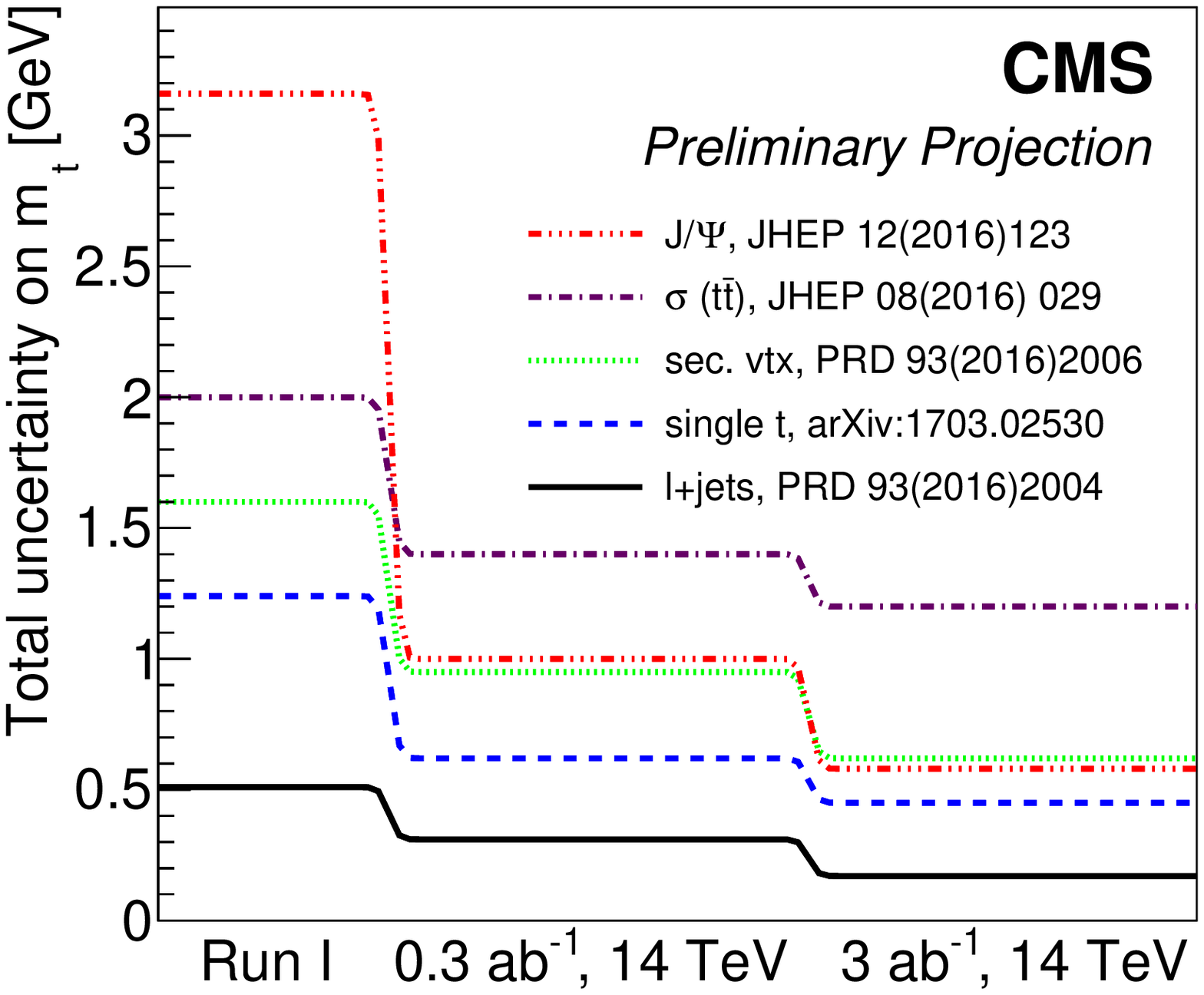}\vspace{-1.5cm}
\caption{Projections of the uncertainty on the top-mass determination
  for different strategies, according to the CMS Collaboration, as a
function of the integrated luminosity at HL-LHC.}
\label{cmshllhc}
\end{figure}

The perspectives for the top-mass determination
at High Luminosity (HL) LHC were debated in
Ref.~\cite{Azzi:2019yne}, where the HL-LHC will collide protons
at 14 TeV and accumulate an integrated luminosity of 3000~fb$^{-1}$.
In the report \cite{Azzi:2019yne} the ATLAS Collaboration presented
a projection for the accuracy on $m_t$ using samples of events
in the lepton+jets mode and $J/\psi\to\mu^+\mu^-$ decays in the
final state, along the lines of Ref.~\cite{ATL-PHYS-PUB-2018-042}.
The expected statistical and systematic uncertanties amount to
0.14 and 0.48 GeV respectively.
As for CMS, the potentials for the top-mass extraction at HL-LHC are 
detailed in \cite{CMS-PAS-FTR-16-006} and summarized in
Fig.~\ref{cmshllhc}: one can learn that all uncertainties
will tremendously decrease at HL-LHC.
In particular, one expects an error which ranges from about 0.2 GeV (0.1\%)
for direct reconstruction in the lepton+jets channel to 1.2 GeV
(0.7\%) from the total $t\bar t$ NNLO cross section.
It is remarkable that the uncertainty from $J/\psi$ final states
will go down to about 0.6 TeV (0.35\%). 

\section{Interpretation of the top-mass measurements
  and theoretical uncertainties}

The nature of the reconstructed top-quark mass and its possible relations
with field-theory mass definitions has lately become
the topic of a very lively debate (see, e.g., the reviews in
\cite{Nason:2017cxd,Hoang:2014oea,Corcella:2015kth}).
I shall first overview the main issues concerning the $m_t$
interpretation and then discuss the dominant sources of theoretical
uncertainty.

\subsection{Measured mass ad theoretical definitions}

The discussion on the identification
of the measured quantity is mostly based on the claim that Monte
Carlo codes are LO, while well-posed field-theory mass definitions
need at least a NLO computation.
Although it is certainly true that, referring to standard codes,
total cross sections are LO, event shapes and differential distributions
go well beyond LO and account for a resummation of enhanced logarithms.
NLO+shower codes like POWHEG and MC@NLO yield
NLO total cross sections, adopting the top pole mass in the computation,
while the differential spectra rely on the shower approximation
and on the modelling of hadronization and underlying event.
Nevertheless, it is indeed cumbersome interpreting the reconstructed top
mass in terms of theoretical definitions or, in other words, scrutinizing
all possible sources of uncertainties which may prevent such
an identification.
As far as this controversy is concerned, one can basically
follow two mainstream viewpoints.

On the one hand, there are
authors \cite{Fleming:2007qr,Hoang:2008xm,Butenschoen:2016lpz,Hoang:2017kmk,Hoang:2018zrp}
who claim that the measured
quantity cannot be directly associated with any field-theory
mass definition and therefore one must stick to the notion 
of Monte Carlo mass. Along this point of view,
much work has been undertaken in order to relate the Monte Carlo
mass to definitions like the pole mass: the quoted
discrepancies between Monte Carlo and pole
masses have through the years ranged from few hundreds MeV to,
in the most extreme case, almost 1 GeV. 
If this were indeed the case, it would be an uncertainty comparable
or even larger than the current errors on the directly reconstructed
top mass.
On the other hand, we have authors \cite{Nason:2016tiy,Nason:2017cxd}
    who instead argue against the use of the Monte
Carlo mass and claim that, under given circumstances, the reconstructed
mass should actually mimic the pole mass. According to this viewpoint,
instead of constructing other mass definitions
to properly interpret the measurements, the effort
should rather be devoted to carefully estimate the
theoretical uncertainties,
of both perturbative and non-perturbative nature, in the
identification of the measured quantity with the pole mass.
In the following, I will briefly review the work carried out in this
respect.

As far as I know, the pioneering work on relating the
measured mass to the pole mass was carried out in
Refs.~\cite{Fleming:2007qr,Hoang:2008xm}. 
First, Ref.~\cite{Fleming:2007qr} defined, for the case study of
$e^+e^-\to t\bar t$ collisions, 
the SCET (MSR-like) short-distance jet mass $m_J(\mu)$,
associated with the collinear jet function and corresponding to the MSR mass
at a scale about the top width, i.e. $R=\Gamma_t$.
Then, $m_J(\mu)$ was related to the pole mass by means of the
following equation:
\begin{equation}\label{jetmass}
m_J(\mu)=m_{\rm pole}-\frac{\alpha_S(\mu)C_F\Gamma_t}{\pi}
\left(\ln\frac{\mu}{\Gamma_t}+\frac{3}{2}\right)+{\mathcal O}(\alpha_S^2).
\end{equation}
Setting, e.g., $\mu\simeq 1$~GeV, then the jet mass differs from the
pole mass by about 200 MeV at ${\mathcal O}(\alpha_S)$. It is also remarkable
that the correction is of order ${\mathcal O}(\alpha_S\Gamma_t)$,
which confirms the intuition that 
the top width has to play 
a role in the uncertainty in the measured mass. 
Later on, Ref.~\cite{Hoang:2008xm} did define a Monte Carlo mass and,
relying on the standard shower implementations, stated that the extracted
top mass could be interpreted as the jet mass evaluated at a scale
of the order of the shower cutoff $Q_0$, i.e. 
$m_t^{\rm MC}\simeq m_t^{\rm MSR}(Q_0)$. Reference~\cite{Hoang:2008xm}
set $Q_0=\left(3^{+6}_{-2}\right)$~GeV and  
$m_t^{\rm MSR}(Q_0)$ to the value of the (Tevatron-based)
top-mass world average at that time, and got
a consistent value of the $\overline{\rm MS}$ mass
$\bar m_t(\bar m_t)$, by using renormalization group evolution
equations.

More recently, Ref.~\cite{Butenschoen:2016lpz} compared PYTHIA
with a SCET computation in the NLO approximation, resumming
soft- and collinear-enhanced contributions to NLL or even NNLL accuracy.
As in \cite{Hoang:2008xm}, the SCET resummed calculation employed the
MSR mass $m_t^{\rm MSR}(R)$, with $R\sim \Gamma_t$ and
$m_t^{\rm MSR}(R)\to m_{t,\rm pole}$ for $R\to 0$.
The PYTHIA mass parameter was then calibrated to reproduce the SCET
prediction for the 2-jettiness
$\tau_2$, after running the code for
several centre-of-mass energies and a few values of the top mass.
The result of Ref.~\cite{Butenschoen:2016lpz} is that the PYTHIA mass 
is consistent, within the errors, with the MSR mass evaluated at a scale
of 1 GeV. Using instead the pole mass in the computation 
yields a shift with respect to the PYTHIA $m_t$ about 600-900 MeV,
according to whether the Monte Carlo results are compared with a NLL
or NNLL resummation.
The work in \cite{Butenschoen:2016lpz}
was extended to $pp$ collisions in Ref.~\cite{Hoang:2017kmk},
where the extraction of $m_t$ from boosted top jets
with light soft-drop grooming 
was proposed.\footnote{Soft-drop grooming is a jet-substructure technique,
  which recursively removes soft wide-angle radiation from a jet.
  See \cite{Larkoski:2014wba} for details.}
By comparing the NLL resummation for the groomed top-jet mass with PYTHIA, the
pole mass was found about 400-700 MeV below the calibrated
Monte Carlo mass, depending on the energy of the $pp$ collision and
non-perturbative parameters contained in the resummation.
Still on this subject, Ref.~\cite{Hoang:2018zrp} explores the
dependence of $m_t$ on the parton shower cutoff, referring
to the HERWIG 7 angular-ordered cascade.
By working in the quasi-collinear limit, with boosted massive quarks
in the NLL approximation, the authors of \cite{Hoang:2018zrp}
stated that the mass parameter in a Monte Carlo code should be
identified with a cutoff-dependent, coherent-branching (CB)  mass, labelled
as $m_t^{\rm CB}(Q_0)$.
Such a coherent-branching mass is
a low-scale short-distance mass, free from renormalon corrections,
related to the pole mass by a relation like:
\begin{equation}\label{mcb}
  m_{t,\rm CB}(Q_0)=m_{t,\rm pole}-\frac{2}{3}\alpha_S(Q_0)Q_0+
  {\mathcal O}(\alpha_S^2Q_0).
  \end{equation}
Expressing in Eq.~(\ref{mcb}) $\alpha_S$ in terms of the 
Monte Carlo QCD scale $\Lambda_{\rm MC}$ 
defined in \cite{Catani:1990rr} and setting $Q_0=1.25$~GeV,
like the shower cutoff of HERWIG 7, 
the shift between pole and CB masses amounts to about 500 MeV.
Using instead the standard $\overline{\rm MS}$
scheme for $\alpha_S$ yields a discrepancy of the order of 300 MeV.
Concerning the calibration of the Monte Carlo mass parameter, 
another approach was suggested in \cite{Kieseler:2015jzh}:
one measures an observable, e.g. a total or differential cross section,
ignoring anything on the event generation, 
and, by comparing the data with the
simulation, calibrates both observable and $m_t$.
The finding of of Ref.~\cite{Kieseler:2015jzh} is that, given the
current precision on the inclusive $t\bar t$ rate, 
the uncertainty on this calibration is roughly 2 GeV.

As anticipated above, 
other authors, such as \cite{Nason:2016tiy,Nason:2017cxd},
claim that it is not really
necessary to introduce the Monte Carlo mass
concept to interpret measurements relying
on final-state direct reconstruction. The starting point 
is the observation
that, in the narrow-width approximation and assuming that one is able
to catch all final-state radiation, the invariant mass of top-decay
products in $t\to bWX$, $X$ being some extra radiation off top and bottom
quarks, should mimic the on-shell top mass, i.e. the pole mass.
Effects due to the top final width, parton emission
which is not included in the reconstruction, contamination from initial-state
radiation and non-perturbative phenomena, such as colour reconnection
or underlying event, clearly spoil the direct identification of the
invariant mass of top-decay final states with the pole mass.
However, in the perspective of Refs.~\cite{Nason:2016tiy,Nason:2017cxd}
rather than a genuine shift of the measured mass with respect to the pole
mass, such effects are seen
as uncertainties, of either perturbative or non-perturbative
nature, in the identification of the extracted mass as pole mass.

Although such approaches may sound pretty different, work towards a
possible compromize was carried out in \cite{Azzi:2019yne}, in such a
way to guide
the top-quark community and avoid confusion or
statements claiming a sort of ignorance on the nature of the measured
top-quark mass.
Though starting from different perspectives,
all those papers agree that the measured $m_t$ can be connected to the
pole mass by means of a relation like:
\begin{equation}\label{ddelta}
  m_t=m_{t,\rm pole}+\delta m_t\pm \Delta m_t,
\end{equation}
where $\delta m_t$ is a possible shift between measured and pole masses
and $\Delta m_t$ is an uncertainty.
According to Refs.~\cite{Nason:2016tiy,Nason:2017cxd},
which basically discourage the use of the concept of Monte Carlo mass,
the extracted mass through top-decay final-state reconstruction
mimics the pole mass, up to some
computable uncertainty. In this approach $\delta m_t\simeq 0$,
while $\Delta m_t$ is a theoretical (Monte Carlo based)
error that, in measurements employing event generators,
should be estimated, e.g., varying shower/hadronization parameters,
confronting different models (cluster and string models for
hadronization are a typical example) or changing the
analysis details (for final-state jets,
increasing/decreasing the jet radius leads to accounting for more or
less gluon radiation). In the view of
Refs.~\cite{Nason:2016tiy,Nason:2017cxd}, the uncertainty
$\Delta m_t$ in the
identification of the measurements with the pole mass should be
of the order of the hadronization scale, i.e.
${\mathcal O}(\Lambda)$.
On the contrary, in the work carried out in
Refs.~\cite{Hoang:2008xm,Butenschoen:2016lpz,Hoang:2017kmk,Hoang:2018zrp}
$m_t$ is labelled as Monte Carlo mass and $\delta m_t$
is an actual discrepancy with respect to the pole mass,
typically about ${\mathcal O}[Q_0\alpha_S(Q_0)]$ as in
Eq.~(\ref{mcb}), while
$\Delta m_t$ is still an uncertainty,
which one can estimate by varying the parameters or options
in the codes and computations employed in the comparison.

Therefore, the disagreement among most authors of the
relevant literature on the interpretation of the top-mass
measurement is conceptually relevant, but in practice concerns 
whether one should calculate an actual discrepancy
$\delta m_t$ or not, as well as
the meaning of $\Delta m_t$ and its numerical magnitude.
In Refs.~\cite{Hoang:2008xm,Butenschoen:2016lpz,Hoang:2017kmk,Hoang:2018zrp}
different values for $\delta m_t$ and $\Delta m_t$ 
have been quoted, which is reasonable,
since, as also advocated in \cite{Nason:2016tiy} for the purpose
of the uncertainty, any possible relation between the
pole mass and the measured quantity has to 
depend on the
observable which is used to extract $m_t$, on the details
of the analysis, such as the imposed cuts, the
energy of the collider
and whether it runs, e.g., $e^+e^-$ or $pp$ modes.
Moreover, since such determinations are based on a comparison between
Monte Carlo results with resummed calculations, with $m_t$ being
a tunable parameter, $\delta m_t$ and $\Delta m_t$
also depend on the accuracy
of the resummations, e.g., NLL or NNLL.
As discussed above, $\delta m_t$ is about 
200 MeV in Ref.~\cite{Hoang:2008xm}, in the range 600-900 MeV
in Ref.~\cite{Butenschoen:2016lpz}, 400-700 MeV in
\cite{Hoang:2017kmk} and
300-500 MeV
in \cite{Hoang:2018zrp}.
The uncertainty $\Delta m_t$ in the relation (\ref{ddelta}) was estimated 
to be roughly
250 MeV in \cite{Hoang:2018zrp} and 280-380 MeV in \cite{Butenschoen:2016lpz}.
Refs.~\cite{Nason:2016tiy,Nason:2017cxd}
  do not contain an explicit calculation of
  $\Delta m_t$, but rather propose a method to compute it, e.g.,
  by varying Monte Carlo perturbative and non-perturbative parameters
  or, in a POWHEG-like implementation,
  switching NLO and width effects on or off.
  Of course, it will be very interesting to follow such an approach
  and compare the results with the numbers obtained
  in Refs.~\cite{Hoang:2008xm,Butenschoen:2016lpz,Hoang:2017kmk,Hoang:2018zrp}.
  One may already guess that, since Refs.~\cite{Nason:2016tiy,Nason:2017cxd}
  do not account for any explicit discrepancy $\delta m_t$, one may
  likely get a larger uncertainty $\Delta m_t$ when following this
  approach.
  Furthermore, it will be crucial understanding how much, for a given
  observable, 
  any shift/uncertainty of the measured mass with respect to the
  pole mass depends on the specific shower code and, e.g., one finds
  an impact of the late implementation of NLO corrections and
  width effects along the lines of \cite{Jezo:2016ujg}.

  \subsection{Theoretical uncertainties in the top mass determination}

For the sake of a precise determination of the top-quark mass,
a reliable estimate of the theoretical error is of paramount importance.
In the top-mass world-average extraction, i.e. Ref.~\cite{ATLAS:2014wva},
based on the so-called standard measurements, 
the overall theory uncertainty accounts for
about 540 MeV of the total 710 MeV systematics.
In particular, Ref.~\cite{ATLAS:2014wva} distinguishes the contributions
due to Monte Carlo generators, radiation effects, colour
reconnection and parton distribution functions (PDFs).

The Monte Carlo systematics is due to the differences in
the implementation of parton showers, matrix-element matching,
width effects, 
hadronization and underlying event in the various programs available
to describe top-quark production and decay.
There is no unique way to
estimate this uncertainty, though, and each collaboration even
follows different prescription according to the analysis.
One can either compare two
different generators, which are considered appropriate for a given
analysis and have been properly tuned to some data sets,
or choose one single code and explore how its
predictions fare with respect to variations of its parameters.
For example, in \cite{ATLAS:2014wva} CDF compares HERWIG and PYTHIA,
while D0 uses ALPGEN+PYTHIA and ALPGEN+HERWIG;
both Tevatron experiments use MC@NLO to gauge the overall
impact of NLO corrections. At the LHC, ATLAS compares
MC@NLO with POWHEG for the NLO contributions and PYTHIA
with HERWIG for shower and hadronization; CMS instead
confronts LO MadGraph with NLO POWHEG.

The radiation uncertainty gauges the effect of
initial- and final-state radiation on the top mass and is typically
obtained by varying in suitable ranges the relevant parameters 
in the parton-shower generators.
Concerning PDFs, there are distinct strategies to evaluate the
induced error on $m_t$ in the different experiments, although
using two different sets or a given set but with different 
parametrizations are common trends.
More generally, the choice of the PDF set in analyses based on
event generators has also been the topic of several discussions:
as pointed out before, although Monte Carlo codes yield LO or
NLO total cross sections, differential spectra go beyond
such approximations and include the resummation of classes of
enhanced logarithmic terms. An attempt to propose some improved
sets of parton distribution functions 
for standard parton shower generators was presented in
\cite{Sherstnev:2007nd}.

Among the sources of theoretical uncertainty
and possibile shifts between measured and pole masses, 
colour reconnection should deserve some special attention.
In fact, it
accounts for about 310 MeV in the world average
presented in \cite{ATLAS:2014wva}.
Also, the very fact that, for example, a bottom 
quark in top decay ($t\to bW$) can be colour-connected to 
an initial-state antiquark does not have its counterpart in
$e^+e^-$ annihilation and therefore its modelling in  
Monte Carlo event generators may need retuning at hadron colliders.
Investigations on the impact of colour reconnection
on $m_t$
were undertaken in \cite{Argyropoulos:2014zoa,Corcella:2014rya},
in the frameworks of PYTHIA and HERWIG, respectively.
In particular, Ref.~\cite{Corcella:2014rya} addresses this
issue by simulating fictitious top-flavoured hadrons $T$ in HERWIG
and comparing final-state distributions, such as the $BW$ invariant
mass, with standard $t\bar t$ events. In fact, in the top-hadron case,
assuming $T$ decays according to the spectator model, the $b$ quark
is forced to connect its colour with the spectator or with antiquarks in its 
own shower, namely $b\to bg$, followed by $g\to q\bar q$,
and colour reconnection is suppressed.
The analysis in \cite{Corcella:2014rya} is still ongoing and, in future
perspectives, it may also serve 
to address the error on the identification of the measured mass
with the pole mass. 
In fact, in the event samples simulated in \cite{Corcella:2014rya}
the Monte Carlo (HERWIG) mass is 
the mass of a heavy hadron, which can be related to
any definition of the
heavy-quark (top for $T$ mesons) mass
definition by means of lattice, potential models or Non Relativistic QCD.
In Ref.~\cite{Argyropoulos:2014zoa}, colour reconnection is instead
investigated within the
Lund string model, tuned
to charged-particle multiplicity or transverse-momentum data.
Several possible models for colour reconnection were investigated and
the yielded uncertainty on the top mass varied between 200 and 500
MeV, depending on the chosen framework.

Another non-perturbative phenomenon which plays a role in the
theoretical error is bottom-quark fragmentation, i.e. the
hadronization of bottom quarks in top decays into $b$-flavoured
mesons or baryons.
The usual way to deal with it consists in tuning the Monte Carlo 
fragmentation parameters to precise $e^+e^-\to b\bar b$ data and 
then using the best parametrizations to describe bottom-quark
hadronization in top decays. This approach was followed, e.g., in
Refs.~\cite{Corcella:2005dk,Corcella:2009rs}, where
data from DELPHI \cite{Abreu:1996na}
SLD \cite{Abe:1999ki}, OPAL \cite{Abbiendi:2002vt}
and ALEPH \cite{Heister:2001jg} were employed to tune the parameters
of HERWIG \cite{Corcella:2000bw} and PYTHIA \cite{Sjostrand:2006za}.
In particular, Ref.~\cite{Corcella:2009rs} used such a tuning
to predict the $B$-hadron+lepton invariant mass $m_{B\ell}$ in
$t\bar t$ events at LHC. A possible extraction of $m_t$ using
this observable exhibited a large discrepancy between
the two event generators, which was explained as due to the
different quality of the $e^+e^-$ fits, with HERWIG being only
marginally consistent with the data.
More recent modelling and fits, such as the 
so-called Monash  \cite{Skands:2014pea} or A14 \cite{a14}, 
or using the dipole-like shower implementation in 
\cite{Cormier:2018tog} are expected to give a better description
of bottom fragmentation in top decays.
Investigations on the uncertainties using these
implementations are currently in progress; it will be very interesting,
in particular, exploring bottom-quark fragmentation by using
NLO+shower codes, such as POWHEG and aMC@NLO, interfaced
to HERWIG or PYTHIA. In fact, it is mandatory to understand whether
the Monte Carlo default parameterizations or tunings like those
in \cite{Skands:2014pea,a14} 
work well at the LHC even when the hard scattering is at NLO,
or one would rather need to refit the Monte Carlo
parameters.
In general, although the approach followed in \cite{Corcella:2009rs}
relies on the universality of the hadronization transition,
it is not absolutely guaranteed that models which reproduce $e^+e^-$ data
work equally well in a coloured environment like
$t\bar t$ events at the LHC, where initial-state radiation,
colour reconnection and underlying event play a role.
Therefore, tuning shower and hadronization parameters
to LHC data should become a ultimate goal.

From this viewpoint, more recently, Ref.~\cite{Corcella:2017rpt}
reconsidered the issue of the dependence of $m_t$ on
Monte Carlo parameters, suggesting
a possible in-situ calibration of the shower codes using top events
in the dilepton channel, 
and taking particular care about observables sensitive to 
$b$-fragmentation in top decays.
In particular, Ref.~\cite{Corcella:2017rpt}
extended the work in  \cite{Corcella:2009rs} exploring
top-decay observables in terms of $B$-hadrons, instead of $b$-jets, so that
one should deal
with fragmentation uncertainties,
rather than with the jet-energy scale.
For instance, if $\langle O\rangle$ is the average value
of a given observable $O$ 
and $\theta$ a generic generator parameter, then
one can write the following relations:
\begin{equation}
  \frac{dm_t}{m_t}=\Delta_O^m\  \frac{d\langle O\rangle}{\langle O\rangle}\ \ ;\ \ \frac{d\langle O\rangle}{\langle O\rangle}=
  \Delta_\theta^O\ \frac{d\theta}{\theta}\  \Rightarrow\ 
  \frac{dm_t}{m_t}=\Delta_\theta^m \frac{d\theta}{\theta},
  \end{equation}
where we defined $\Delta_\theta^m=\Delta_O^m\ \Delta_\theta^O$.
Therefore, if one aims at, e.g., an error of 500 MeV on 
$m_t$, namely $dm_t/dm_t<0.003$, one should also have
$\Delta_\theta^m (d\theta/\theta)<0.003$.
Reference~\cite{Corcella:2017rpt} then identifies some
so-called calibration observables, which depend on the shower/hadronization
parameters but are rather insensitive to the top mass.
Examples of such quantities are, e.g., the ratios of
$B$-hadron to $b$-jet ($b$) transverse momenta $p_{T,B}/p_{T,b}$, of 
invariant masses $m_{B\bar B}/m_{b\bar b}$ ($\bar b$ being
a jet containing a $\bar B$ hadron), the azimuthal separations
and invariant opening angles
$\Delta\phi(b\bar b)$, $\Delta\phi(B\bar B)$, $\Delta R(b\bar b)$,
$\Delta R (B\bar B)$.\footnote{For two particles at 
  $\Delta\phi$ and $\Delta\eta$ distances in azimuth and rapidity,
  the invariant opening angle is defined as
  $\Delta R=\sqrt{(\Delta\phi)^2+(\Delta\eta)^2}$.}
Then, imagining that one could ideally 
tune the parameters to measurements of the
calibration observables, other quantities can be explored to
extract $m_t$, such as the $B$-hadron energy and transverse momentum
$E_B$ and $p_{T,B}$, or the invariant masses $m_{B\ell}$,
$m_{\ell\bar\ell}$ and $m_{B\bar B\ell\bar \ell}$.
The conclusion of this exploration is that, in order to achieve a
0.3\% precision on the top mass, one needs to determine
the strong coupling constant at 1\% accuracy and other parameters,
such as the shower cutoff, the gluon and quark
effective masses or the hadronization parameters at 10\%.
Overall, Ref.~\cite{Corcella:2017rpt} proposes a method to tune directly 
Monte Carlo generators to data from top events at the LHC, which,
whenever top-production data were to become precise enough, 
should be preferable to the use of fits to $e^+e^-$ data, in such
a way to avoid all uncertainties and ambiguities in the
application of $e^+e^-$-based fits to hadron collisions.

\section{Conclusions}

I discussed some challenging issues regarding the determination
and interpretation of the top
quark mass at hadron colliders. I reviewed the main top mass definitions,
pointing out their most notable features
and taking particular care about the
pole and $\overline{\rm MS}$ masses. 
I described recent calculations for the purpose of 
the renormalon ambiguity in the pole mass in the
infrared regime, yielding uncertainties about 100-250 MeV, which, for the
time being, are below the current error on the top mass.
Also, such estimates are well below the top-width energy
scale, about 1.4 GeV.

The most relevant features of Monte Carlo codes for top-quark
phenomenology were then presented, stressing the late implementation
of NLO corrections and interference effects between 
top-production and decay phases.
Even the standard shower codes are nevertheless beyond LO in the
differential distributions which account for classes of
enhanced soft/collnear logarithms to all orders.

The main experimental methods to measure the top mass were discussed,
pointing out the differences among the so-called standard
and alternative measurements and the magnitude of the quoted uncertainties.
For the time being, although the alternative measurements provide
an excellent ground to reconstruct $m_t$ using the kinematic
properties of the final states and, in some cases, they are even
capable of minimizing the impact of the chosen Monte Carlo generator,
the standard methods are still those which yield the lowest
uncertainty. This will also be the case in the future LHC runs,
albeit the higher statistics are expected to decrease the
errors in the alternative strategies too.

Much space was then 
devoted to the present debate on the interpretation
of the measurements and whether one should relate the extracted $m_t$ to some
alternative mass definition or rather express it in terms of the
pole mass, up to some uncertainty.
A common features of both attitudes is nonetheless that there is no
universal relation between the measured mass and any field-theory
definition, but it depends on the considered observable and on the
type of Monte Carlo shower code or QCD calculation which is employed
in the comparison. There have been many investigations to relate
the measured $m_t$ to short-distance masses by comparing Monte Carlo
predictions with SCET resummed computations: the obtained shift
with respect to the pole mass was eventually derived and
is of the order of a few hundreds of MeV, depending on the
specific analysis and accuracy of the calculation.
On the other hand, work is in progress to explore the sources
of errors which, on the top of the theoretical systematics, 
affect the straightforward identification of the top mass in 
direct-reconstruction analyses as a pole mass,
such as colour reconnection.
Although the starting point of such approaches are conceptually
different, a compromize can be reached and it will be very 
appealing applying the ongoing work on colour-reconnection and
bottom-fragmentation uncertainties to
the interpretation of the top-mass measurements in terms of
well-defined field-theory quantities.

Finally, referring in particular to the world-average analysis,
the contributions to the quoted theoretical error were debated,
along with the current work aimed at obtaining even
more reliable estimates of such uncertainties.
Furthermore, it was discussed
the possibility to use top-quark events and suitable
calibration observables to fit Monte Carlo parameters,
which will probably
be the way to follow in future perspectives, once the data become
precise enough to compete with $e^+e^-$ experiments for the
purpose of the tuning of event generators.

In summary, top-quark phenomenology at the LHC, especially in the
high-luminosity perspective, has become precision physics and
the smallness of the current and foreseen uncertainties in the top-mass
measurement are a clear example of such a level of accuracy.
However, for the sake of a robust and reliable top-mass determination,
much work is still necessary, in order to understand better and possibly
reduce the sources of uncertainties. In particular, progress in
Monte Carlo studies and QCD calculations for top production and decay, 
as well as in theoretical work concerning top-mass definitions,
should definitely be encouraged.
As pointed out many times in this review, investigations along
these lines are already in progress, in such a way that one can feel
confident that the theoretical and experimental efforts will eventually
converge to match the precisions which are expected in the future LHC runs
and ultimately at HL-LHC.

\section*{Acknowledgments}
I acknowledge discussions with Andr\'e Hoang and Paolo Nason,
co-authors of the top-mass section in Ref.~\cite{Azzi:2019yne}, on
the interpretation of the top-mass measurements.

\bibliography{mt2}{}

\end{document}